\documentclass[aps,prd,groupedaddress]{revtex4}

\usepackage{amsfonts}
\usepackage{graphicx}

\begin{document}


\title{Strongly coupled quantum field theory}


\author{Marco Frasca}
\email[]{marcofrasca@mclink.it}
\affiliation{Via Erasmo Gattamelata, 3 \\ 00176 Roma (Italy)}


\date{\today}

\begin{abstract}
We analyze numerically a two-dimensional $\lambda\phi^4$ theory showing that in the
limit of a strong coupling $\lambda\rightarrow\infty$ just the homogeneous solutions 
for time evolution are relevant in
agreement with the duality principle in perturbation theory as presented in
[M.Frasca, Phys. Rev. A {\bf 58}, 3439 (1998)], being negligible the contribution of
the spatial varying parts of the dynamical equations. 
A consequence is that the Green function method
works for this non-linear problem in the large coupling limit as in 
a linear theory. A numerical proof is given for this. With these results at hand,
we built a strongly coupled quantum field theory for a $\lambda\phi^4$ interacting field
computing the first order correction to the generating functional. Mass spectrum
of the theory is obtained turning out to be that of a harmonic oscillator with
no dependence on the dimensionality of spacetime. The agreement with the Lehmann-K\"allen
representation of the perturbation series is then shown at the first order.
\end{abstract}


\maketitle


A lot of problems in physics have such a difficult equations to solve that the most natural
approach is a numerical one. Weak perturbation theory generally proves to be insufficient
to extract all the physics. A well-known case is given by quantum chromodynamics that due to
the strength of the coupling constant at low energies, makes useless known perturbation
techniques demanding the need for numerical solutions.

In the seventies and eighties of the last century 
a significant attempt to build a perturbation theory
for a strongly interacting quantum field theory was proposed \cite{kov,pmb,be1,par,be2,be3,coo,be4}.
In this approach it was stipulated that the perturbation to be considered is the
free part of the Lagrangian. 
Notwithstanding this approach is still studied today \cite{svai} 
no fruitful results have been obtained so far due to the strongly singular perturbation series
that is obtained in this way. Rather, the rationale behind this method is really smart as one
recognize that just interchanging the two parts of the Lagrangian one gets different perturbation
series. 

This duality in perturbation theory is a general mathematical property of differential
equations as was shown in Ref.\cite{fra1,fra2}. What makes duality interesting is
the general property of the leading order that, while in the weak perturbation case is just a
free linear theory whose solution is generally known, for the dual series that holds in
the limit of a strongly coupling, that is a coupling going to infinity, one can prove a
theorem showing that the adiabatic approximation applies. 
We also pointed out in recent works \cite{fra3,fra4} that in field theory and general relativity
the dual perturbation series at the leading order produces a rather interesting result: in
a strongly coupled field theory the leading order is ruled by a homogeneous equation, that is,
the spatial variation of the field in the equations of the theory becomes negligible. In
general relativity this gives precious informations on the space-time
near a singularity where the above behavior was conjectured in \cite{lk,blk1,blk2} and
numerically shown in \cite{garf}.

In this paper we have two different aims. Firstly, we intend to prove that the
numerically observed behavior in general relativity also holds for a $\lambda\phi^4$ theory, that
is, a homogeneous equation rules the leading order of a strongly interacting scalar field. 
Then, after numerically proving that in a strongly coupled field theory the Green
function can be used in the same way as done in a weak field theory, a quantum
field theory is obtained.

We apply the duality principle in perturbation theory as devised in \cite{fra1,fra2,fra3,fra4}
assuming a Hamiltonian for the field (here and in the following we take $\hbar=c=1$)
\begin{equation}
    H = \int d^{D-1}x\left[\frac{1}{2}\pi^2+\frac{1}{2}(\partial_x\phi)^2+
	\frac{1}{2}\mu_0^2\phi^2+
	\lambda V(\phi)\right]
\end{equation} 
being $D$ the dimension, $\mu_0$ the mass and $\lambda$ the coupling. For our aims we take $\mu_0=1$
and a single component scalar field. This Hamiltonian gives the following Hamilton equations
\begin{eqnarray}
\label{eq:phi}
    \partial_t\phi &=& \pi \\ \nonumber
	\partial_t\pi  &=& \partial_x^2\phi -\phi -\lambda V'(\phi)
\end{eqnarray}
apex meaning derivation with respect to $\phi$.
From eqs.(\ref{eq:phi}) we recognize two perturbation terms 
$\partial_x^2\phi -\phi$ and $V'(\phi)$ and one may ask what is the relation between
the weak perturbation series for the latter term with
the one having the term $\partial_x^2\phi -\phi$ as a perturbation.
Indeed, by exchanging $\partial_x^2\phi -\phi\leftrightarrow V'(\phi)$ 
for perturbation the following equations can be obtained
\begin{eqnarray}
    \partial_{\tau}\phi_0 &=& \pi_0 \\ \nonumber
	\partial_{\tau}\phi_1 &=& \pi_1 \\ \nonumber
	                 &\vdots&  \\ \nonumber
	\partial_{\tau}\pi_0 &=& -V'(\phi_0) \\ \nonumber
	\partial_{\tau}\pi_1 &=& -V''(\phi_0)\phi_1 + \partial_x^2\phi_0-\phi_0 \\ \nonumber
	                 &\vdots&.
\end{eqnarray}
This is a non trivial set of equations that can be recovered if we take
\begin{eqnarray}
   \tau &=& \sqrt{\lambda}t \\ \nonumber
   \pi &=& \sqrt{\lambda}\left(\pi_0 + \frac{1}{\lambda}\pi_1 + \frac{1}{\lambda^2}\pi_2 + \ldots\right) \\ \nonumber
   \phi &=& \phi_0 + \frac{1}{\lambda}\phi_1 + \frac{1}{\lambda^2}\phi_2 + \ldots.
\end{eqnarray}
So, our interchange of the perturbations produced a dual series that holds in the limit $\lambda\rightarrow\infty$
as expected by the duality principle in perturbation theory \cite{fra1,fra2}.
The most important result we have obtained is that we get at the leading order a
homogeneous equation, that is, {\sl a self-interacting scalar 
field with a coupling constant going to infinity is ruled by a homogeneous equation}.
This result is relevant as settles the
physical meaning of homogeneous solutions for a given field theory. 

Now, let us specialize the above analysis to a $\lambda\phi^4$ theory. When $\lambda\rightarrow\infty$
we have to solve the leading order equation $\partial_{t}^2\phi_0 = -\lambda\phi_0^3$
that has the following solution by Jacobi elliptic functions \cite{gr}
\begin{equation}
\label{eq:sol}
    \phi_0=\left(2C_1\right)^{\frac{1}{4}}
	{\rm sn}\left[\left(\frac{C_1}{2}\right)^{\frac{1}{4}}(\sqrt{\lambda}t+C_2),i\right]
\end{equation}
being ${\rm sn}$ the snoidal Jacobi elliptic function, $C_1$ and $C_2$ two integration constants
that can depend on spatial variables. So, this analytical solution has
to coincide with the numerical solution of the equation $\Box\phi + \phi + \lambda\phi^3= 0$ with $\lambda$ very
large, after the proper boundary conditions are set.
Another interesting problem is to see how farther can be considered to hold the approximation
\begin{equation}
\label{eq:ma}
    \phi_0(x,t)=\int_0^{\infty}G(t-t')j(x,t')dt'
\end{equation}
as a solution of the equation $\Box\phi + \phi + \lambda\phi^3= j$ in the limit of
$\lambda$ very large and $G(t-t')$ the Green function given by the equation
$\partial_{t}^2 G(t) + \lambda G^3(t)=\delta(t)$ that is
\begin{equation}
\label{eq:gf}
    G_+(t)=\theta(t)\left(\frac{2}{\lambda}\right)^{\frac{1}{4}}
	{\rm sn}\left[\left(\frac{\lambda}{2}\right)^{\frac{1}{4}}t,i\right].
\end{equation}
The time reversed solution
\begin{equation}
\label{eq:gfr}
    G_-(t)=-\theta(-t)\left(\frac{2}{\lambda}\right)^{\frac{1}{4}}
	{\rm sn}\left[\left(\frac{\lambda}{2}\right)^{\frac{1}{4}}t,i\right]
\end{equation}
also holds. It is not difficult to verify that $G_-(t)=G_+(-t)$.

\begin{figure}[tbp]
\begin{center}
\includegraphics[angle=-90,width=240pt]{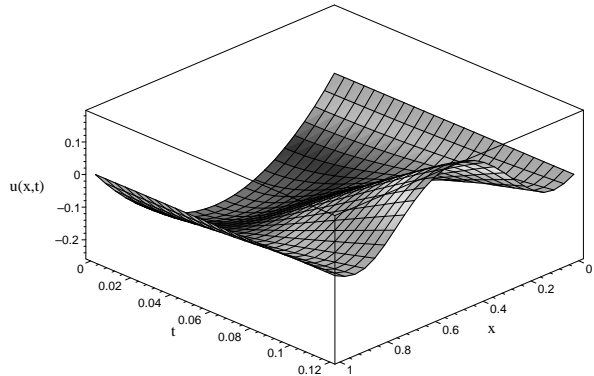}
\caption{\label{fig:fig1} Numerical solution for $\Box\phi + \phi + \lambda\phi^3=0$
with $\lambda=10^4$.}
\end{center}
\end{figure}

\begin{figure}[tbp]
\begin{center}
\includegraphics[angle=-90,width=240pt]{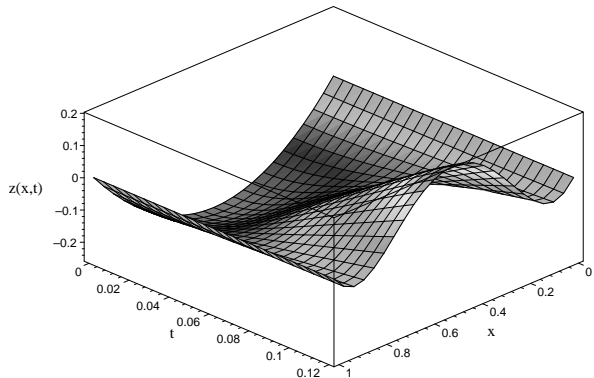}
\caption{\label{fig:fig2} Analytical solution of $\partial_t^2\phi + \lambda\phi^3=0$
with $\lambda=10^4$ as given in eq.(\ref{eq:ana}).}
\end{center}
\end{figure}

The first numerical analysis we worked out is to verify that indeed, when $\lambda$ is very large,
a good first approximation is given by the leading order solution (\ref{eq:sol}). In order
to check this we consider the equation $\Box\phi + \phi + \lambda\phi^3=0$ for $D=2$,
$\lambda=10^4$ and take $\phi(0,t)=0$, $\phi(1,t)=0$, $\partial_t\phi(x,0)=0$ and
$\phi(x,0)=x^2-x$. The solution is given in fig. \ref{fig:fig1}.
The analytical solution in this case can be easily computed by eq.(\ref{eq:sol}) giving
\begin{equation}
\label{eq:ana}
    \phi\approx(x^2-x){\rm sn}\left[(x^2-x)\sqrt{\frac{\lambda}{2}}t+x_0,i\right].
\end{equation}
being $x_0={\rm cn}^{-1}(0,i)$ as to have ${\rm sn}(x_0,i)=1$. 
This solution is presented in fig. \ref{fig:fig2} and the
comparison with the numerical result is quite satisfactory.
Homogeneous solutions drive, in a first approximation, strongly self-interacting
scalar fields.

\begin{figure}[tbp]
\begin{center}
\includegraphics[angle=-90,width=240pt]{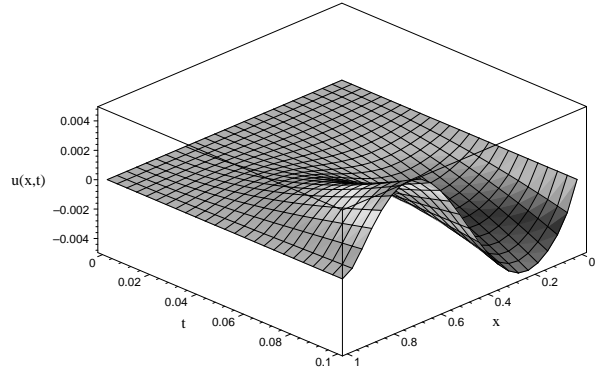}
\caption{\label{fig:fig4} Numerical solution for $\Box\phi + \phi + \lambda\phi^3=-\sin(2\pi(x+t))$
with $\lambda=10^4$.}
\end{center}
\end{figure}

\begin{figure}[tbp]
\begin{center}
\includegraphics[angle=-90,width=240pt]{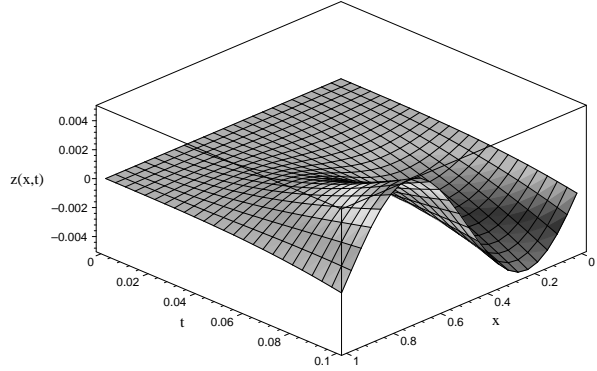}
\caption{\label{fig:fig5} Analytical solution with the Green function of eq.(\ref{eq:gf}) 
and $\lambda=10^4$ and forcing function $j=-\sin(2\pi(x+t))$.}
\end{center}
\end{figure}

For the next step we have studied the $D=2$ equation $\Box\phi + \phi + \lambda\phi^3=-\sin(2\pi(x+t))$
with the same value for $\lambda$ with boundary conditions 
$\phi(0,t)=0$, $\phi(1,t)=0$, $\partial_t\phi(x,0)=0$ and $\phi(x,0)=0$.
The numerical solution is given in fig. \ref{fig:fig4}
The analytical solution can be easily computed with the Green function of eq.(\ref{eq:gf})
giving $\phi\approx-\int_0^tG_+(t-t')\sin(2\pi(x+t'))dt'$
and the result is given in fig.\ref{fig:fig5} and again is quite satisfactory.
These results support the other conclusion that {\sl the Green function method is still
useful in a regime of largely coupled scalar fields}. 

There is an exterminate literature for quantum field theory (see e.g. \cite{zj,iz,ra} for scalar fields).
As a convention we will use
boldface for $D-1$ dimensional vectors. Spacetime signature is $(+,-,-,-)$.
We start with the standard path integral formulation for the generating functional as
$Z[j]=\int[d\phi]e^{\left\{i\int d^Dx\left[
\frac{1}{2}(\partial_t\phi)^2-\frac{1}{2}(\nabla\phi)^2-\frac{1}{2}\phi^2-\lambda V(\phi)+j\phi
\right]\right\}}$
that we rewrite as
\begin{equation}
\label{eq:z}
   Z[j]=\int[d\phi]e^{\left\{i\int d^Dx\left[
   \frac{1}{2}(\partial_t\phi)^2-\lambda V(\phi)+j\phi
   \right]\right\}}
   e^{\left\{-i\int d^Dx\left[\frac{1}{2}(\nabla\phi)^2+\frac{1}{2}\phi^2
   \right]\right\}}
\end{equation}
separating the leading term from the perturbation in agreement with our discussion above. Using our
conclusions about Green function obtained above one can write down the generating
functional, without the perturbation, by the Gaussian approximation
\begin{equation}
\label{eq:z0}
    Z_0[j]=\exp\left[\frac{i}{2}\int d^Dx_1d^Dx_2j(x_1)\Delta(x_1-x_2)j(x_2)\right]
\end{equation}
from which one can get the Wick theorem.
It is easy to verify that $\left.\frac{\delta^2Z_0[j]}{\delta j(y_1)\delta j(y_2)}\right|_{j=0}=
-i\Delta(y_2-y_1)$ having set
\begin{equation}
\label{eq:fp}
    \Delta(x_2-x_1)=\delta^{D-1}(x_2-x_1)[G_+(t_2-t_1)+G_-(t_2-t_1)]=\Delta(x_1-x_2).
\end{equation}
 
In order to make all the argument self-consistent we derive the generating functional (\ref{eq:z0})
from eq.(\ref{eq:z}). The existence of the the leading order functional will rely in the
end on the existence of the semiclassical approximation for the path integral
\begin{equation}
\label{eq:lead}
   Z_0[j]=\int[d\phi]e^{\left\{i\int d^Dx\left[
   \frac{1}{2}(\partial_t\phi)^2-\lambda V(\phi)+j\phi
   \right]\right\}}.
\end{equation}
This can be seen in the following way. Let us apply the rescaling of time $\tau=\sqrt{\lambda}t$.
One has
\begin{equation}
   Z_0[j]=\int[d\phi]e^{\left\{i\sqrt{\lambda}\int d^Dx\left[
   \frac{1}{2}(\partial_{\tau}\phi)^2-V(\phi)
   \right]\right\}}
   e^{\frac{i}{\sqrt{\lambda}}\int d^Dx j\phi}
\end{equation}
that shows that the limit $\lambda\rightarrow\infty$ corresponds to the semiclassical limit.
That is, the system tends to recover a classical behavior in the strong coupling limit and all
the results obtained above for this case apply. So, we reinsert the original time variable $t$
and take
\begin{equation}
\label{eq:phic}
    \phi = \phi_c +\delta\phi
\end{equation}
being $\delta\phi$ a small deviation from the classical solution $\phi_c$ that satisfies the equation
\begin{equation}
\label{eq:mot}
    \ddot\phi_c+\lambda V'(\phi_c)=j.
\end{equation}
Inserting eq.(\ref{eq:phic}) into the functional integral (\ref{eq:lead}) one has, using
eq.(\ref{eq:mot}),
\begin{equation}
     Z_0[j]=   e^{\frac{i}{2}\int d^Dx j\phi_c}F[\phi_c].
\end{equation}
being
\begin{equation}
     F[\phi_c]=e^{-i\lambda\int d^Dx[V(\phi_c)-\frac{1}{2}\phi_cV'(\phi_c)]}
	 \int[d\delta\phi]e^{-i\int d^Dx\left[\frac{1}{2}\delta\phi\ddot\delta\phi
	 +\frac{1}{2}\lambda V''(\phi_c)(\delta\phi)^2
	 \right]}.
\end{equation}
We now apply the property that we have proved for the solution of eq.(\ref{eq:mot}), that is,
the Green function method still applies as in eq.(\ref{eq:ma}). This gives back the
Gaussian functional (\ref{eq:z0}) taking into account that, in the limit of interest
$\lambda\rightarrow\infty$, after the substitution of the Green function (\ref{eq:gf})
and (\ref{eq:gfr}), $F[\phi_c]\approx 1$.

A short digression on the Feynman propagator (\ref{eq:fp}) is needed. Indeed, it is well known that \cite{gr}
\begin{equation}
    {\rm sn}(u,i)=\frac{2\pi}{K(i)}\sum_{n=0}^\infty\frac{(-1)^ne^{-(n+\frac{1}{2})\pi}}{1+e^{-(2n+1)\pi}}
    \sin\left[(2n+1)\frac{\pi u}{2K(i)}\right]
\end{equation}
being $K(i)=\int_0^{\frac{\pi}{2}}\frac{d\theta}{\sqrt{1+\sin\theta}}\approx 1.3111028777$ a constant.
This means that a Fourier transform gives
\begin{equation}
    \Delta(\omega)=\sum_{n=0}^\infty\frac{B_n}{\omega^2-\omega_n^2+i\epsilon}
\end{equation}
being $B_n=(2n+1)\frac{\pi^2}{K^2(i)}\frac{(-1)^{n+1}e^{-(n+\frac{1}{2})\pi}}{1+e^{-(2n+1)\pi}}$
and the mass spectrum of the theory in the limit $\lambda\rightarrow\infty$ is given by
$\omega_n = \left(n+\frac{1}{2}\right)\frac{\pi}{K(i)}\left(\frac{\lambda}{2}\right)^{\frac{1}{4}}$
that we can recognize as those of a harmonic oscillator. 
A mass gap computed for $n=0$ is given by $\delta_S=
\frac{\pi}{2K(i)}\left(\frac{\lambda}{2}\right)^{\frac{1}{4}}$.
This result does not depend on the dimension
$D$ but could depend on the number of components of the scalar field that we have not considered here. 

It is straightforward to write down the full generating functional to work out perturbation theory. One has
\begin{equation}
    Z[j]=\exp\left[\frac{i}{2}\int d^Dy_1d^Dy_2\frac{\delta}{\delta j(y_1)}(-\nabla^2+1)\delta^D(y_1-y_2)
    \frac{\delta}{\delta j(y_2)}\right]Z_0[j]
\end{equation}
that, by expanding the first exponential, gives
\begin{eqnarray}
    Z[j]&=&\left\{1-\frac{1}{2}\int d^Dy_1d^Dy_2(-\nabla^2+1)\delta^D(y_1-y_2)\Delta(y_1-y_2)\right. \\ \nonumber
	&-&\left.\frac{i}{2}\int d^Dy_1d^Dy_2(-\nabla^2+1)\delta^D(y_1-y_2)I(y_1)I(y_2)+\ldots\right\}Z_0[j]
\end{eqnarray}
being $I(z)=\int d^Dx_1\Delta(z-x_1)j(x_1)$.
We realize straightforwardly that there seems to be a divergence as also happens for weak perturbation theory.
In order to make the computation physically clear we pass to momentum space by a Fourier transform as
$\tilde f(k)=\int d^Dx f(x)e^{ikx}$
and one has straightforwardly
\begin{equation}
    Z_0[j]=\exp\left[\frac{i}{2}\int\frac{d^Dk}{(2\pi)^D}\tilde j(k)\tilde\Delta(k)\tilde j(-k)\right].
\end{equation}
So the first integral becomes
\begin{equation}
    \int d^Dy_1d^Dy_2(-\nabla^2+1)\delta^D(y_1-y_2)\Delta(y_1-y_2)=
	\int d^Dkd^Dk_1({\bf k}^2+1)\tilde\Delta(k)\delta^D(k+k_1)\delta^D(k+k_1)
\end{equation}
and we can dispose of the product of Dirac distributions by substituting one of them with
the $D$-dimensional volume $V_D$ divided by $(2\pi)^D$ reducing it to
$V_D \int\frac{d^{D-1}k}{(2\pi)^{D-1}}({\bf k}^2+1)\int\frac{d\omega}{2\pi}\tilde\Delta(\omega)$
where we have explicitly given the dependence on $\omega$ to make clear that this integral
seems to diverge and a cut-off in $k$ has to be introduced.
But we notice that the last integral is nothing else than $\Delta(0)=0$
and so, we take this renormalization constant to be zero. So, finally one has
\begin{equation}
    Z[j]=\left[1-\frac{i}{2}\int\frac{d^Dk}{(2\pi)^D}({\bf k}^2+1)
	\tilde j(k)\tilde\Delta(k)\tilde j(-k)\tilde\Delta(-k)+\ldots
	\right]Z_0[j]
\end{equation}
that is the result we aimed to. We notice that to recover the proper ordering in $\lambda$
one has to turn back to space and time variables and one can see that we are at order
$\lambda^{-\frac{1}{2}}$ having the product of two Green functions. We have an expansion
that holds in the strong coupling limit as promised. We see that this series recover
the proper dependence on ${\bf k}$ in the propagator in agreement with the Lehmann-K\"allen
representation \cite{zj,iz}. This completes the proof of existence of a strongly coupled
quantum field theory for a $\lambda\phi^4$ model.

Recently it was shown by Kleinert as very fine results for critical exponents can be obtained
with the variational method \cite{kle1,kle2,kle3} but no hint is given on the structure of
the solution of the field equations. Here we have built a successful approach showing
a possible way to find solutions to non-linear quantum field theories in the strong
coupling limit.
We were also able to show that a homogeneous equation rules the dynamics and Green
function methods can be successfully applied in the strong coupling limit. All this
has been supported by numerical results. So, this approach can open up the way to exploit
analytical solutions where, presently, just heavy numerical work can be accomplished.



\begin{thebibliography}{99}

\bibitem{kov} S. Kovesi-Domokos, Nuovo Cim. A {\bf 33}, 769 (1976).
\bibitem{pmb} R. Benzi, G. Martinelli, and G. Parisi, Nucl. Phys. B {\bf 135}, 429 (1978).
\bibitem{be1} C. M. Bender, F. Cooper, G. S. Guralnik, and D. H. 
Sharp, Phys. Rev. D {\bf 19},
1865 (1979).
\bibitem{par} N. Parga, D. Toussaint, and J. R. Fulco, Phys. Rev. D {\bf 20},
887 (1979).
\bibitem{be2} C. M. Bender, F. Cooper, G. S. Guralnik, D. H. Sharp, R. Roskies, and 
M. L. Silverstein, Phys. Rev. D {\bf 20}, 1374 (1979).
\bibitem{be3} C. M. Bender, F. Cooper, G. S. Guralnik,
H. Moreno, R. Roskies, and D. H. Sharp, Phys. Rev. Lett. {\bf 45}, 501 (1980).
\bibitem{coo} F. Cooper, and R. Kenway, Phys. Rev. D {\bf 24}, 2706 (1981).
\bibitem{be4} C. Bender, F. Cooper, R. Kenway, and L. M. Simmons, Phys. Rev. D
{\bf 24}, 2693 (1981).
\bibitem{svai} N. F. Svaiter, Physica A {\bf 345}, 517 (2005).
\bibitem{fra1} M. Frasca, Phys. Rev. A {\bf 58}, 3439 (1998).
\bibitem{fra2} M. Frasca, Phys. Rev. A {\bf 60}, 573 (1999).
\bibitem{fra3} M. Frasca, hep-th/0508246.
\bibitem{fra4} M. Frasca, hep-th/0509125.
\bibitem{lk} I. M. Kalathnikov, and E. M. Lifshitz, Phys. Rev. Lett. 24, 76 (1970).
\bibitem{blk1} V. A. Belinski, I. M. Kalathnikov, and E. M. Lifshitz, Adv. Phys. {\bf 19}, 525 (1970).
\bibitem{blk2} V. A. Belinski, I. M. Kalathnikov, and E. M. Lifshitz, Adv. Phys. {\bf 31}, 639 (1982).
\bibitem{garf} D. Garfinkle, Phys. Rev. Lett. {\bf 93}, 161101 (2004).
\bibitem{gr} I. S. Gradshteyn, I. M. Ryzhik, {\sl Table of Integrals, Series, and Products},
(Academic Press, 2000). 
\bibitem{zj} J. Zinn-Justin, {\sl Quantum Field Theory and Critical Phenomena}, (Clarendon Press, 1996).
\bibitem{iz} C. Itzykson, J.-B. Zuber, {\sl Quantum Field Theory}, (McGraw-Hill, 1980).
\bibitem{ra} P. Ramond, {\sl Field Theory: A Modern Primer}, (Addison-Wesley, 1989).
\bibitem{kle1} H. Kleinert, Phys. Rev. D {\bf 57}, 2264 (1998).
\bibitem{kle2} H. Kleinert, Phys. Rev. D {\bf 58}, 107702 (1998).
\bibitem{kle3} H. Kleinert, Phys. Rev. D {\bf 60}, 085001 (1999).
\end{thebibliography}
\end{document}